\documentclass[12pt]{elsarticle}  
\usepackage{amsmath,color,amssymb,hyperref,graphics}
\definecolor{darkblue}{rgb}{0.0,0.0,0.3}
\hypersetup{colorlinks,breaklinks,
            linkcolor=darkblue,urlcolor=darkblue,
            anchorcolor=darkblue,citecolor=darkblue}
\journal{Mathematics and Computers in Simulation}
        
\begin{document}
\begin{frontmatter}

\title{Topological and Geometrical Random Walks on Bidisperse Random Sphere
Packings}
\author{Peter Hinow}
\address{Department of Mathematical Sciences, University of
Wisconsin - Milwaukee, P.O. Box 413, Milwaukee, WI 53201-0413, USA; email:
\texttt{hinow@uwm.edu}}
\date{\today}

\begin{abstract}
Motivated by the problem of predicting the release kinetics of matrix tablets,
we study random walks on the contact graph of a random sphere packing of
spheres of two sizes. For a random walk on the unweighted graph that terminates
in a specified target set, we compare the euclidean distance covered to the
number of steps. We find a linear dependence of the former on the latter, with
proportionality constant the edge length expectation of the contact graph. This
result makes it possible to compare predictions of diffusion path lengths on
geometric graphs. 
\end{abstract}

\end{frontmatter}

\section{Introduction}\label{section:Introduction} 
Diffusion processes in disordered and porous media are ubiquitous in natural and
technical sciences. The diffusion of ground water contaminants in soils
\cite{Bear} or the diffusion of water in the grey  and white matter of the brain
(diffusion magnetic resonance imaging) are only two examples. Oftentimes,
experiments such as the observation of tracer particles are complemented by
numerical simulations. These simulations can use models that are either discrete
or continuous with respect to space and/or time. A special discrete simulation
strategy is random walk on graphs. The graph is chosen to resemble some
properties of the surrounding environment such as 
porosity or tortuosity. At each discrete time step the random walker selects the
position to be occupied after the next step. In the simplest case all adjacent
vertices are selected with equal probability. The procedure terminates once a
certain specified target set has been reached. The number of steps since the
beginning of the walk can then be used to estimate the diffusion path length.

Recently, we developed a mathematical model to predict the release kinetics of
matrix tablets \cite{DCDS_B}. A matrix tablet is a device used to deliver a
pharmaceutically active drug and to release it in a controlled fashion over an
extended period of time, longer than if an immediate release tablet were used.
Matrix tablets are formulated as powder mixtures of the pharmaceutically active
drug and at least two inactive ingredients, namely water soluble excipients and
water insoluble polymers. Upon placement of the compressed tablet in a fluid,
the
polymer matrix remains largely intact while soluble excipient and drug particles
are dissolved and carried away by diffusion. It is of great interest to predict
the time course of the drug release and its dependence on the composition of the
powder mixture as well as compaction pressure and possible curing temperature.
In \cite{DCDS_B} we proposed a mathematical model for the drug
release process, based on a discrete random walk model. This work serves as the
starting point for the present paper.

In \cite{DCDS_B}, we begin with the creation of a random dense sphere packing
$\mathcal{P}$. Random dense sphere packings have found a large number of
applications in a variety of fields, see e.g.~\cite{TorquatoBook,
Knott}. To the best of our knowledge, a mathematically rigorous definition of
this notion is still missing, see \cite{Torquato,Biazzo,Kamien,Song,Hermes,Farr}
and references therein for some recent discussions. Here, we will work with the
result of a procedure known as the Lubachevsky-Stillinger (LS) protocol
\cite{Lubachevsky}. Briefly, a fixed number of spheres move within a container.
At the same time their radii increase at a common rate (which is not necessarily
constant with respect to time). The spheres collide elastically with each other
and with the walls of the container if such are present. Since all spheres grow
at the same rate, the ratios between the radii are preserved. This allows
creation of sphere packings with multiple radius classes, in particular
bidisperse packings when spheres are of one of two sizes. The procedure is
stopped once the collision frequency or alternatively the pressure (the sum of
the squared velocities), exceeds a certain threshold set by the user. From the
random sphere packing so obtained, we construct the contact graph by putting an
edge between sphere centers that are within a certain distance of each other. In
our implementation in \cite{DCDS_B} we label some particles as ``drug'' and
``polymer'' particles according to the formulation of the powder mixture. We
seek their escape routes to ``exterior'' particles with the help of Monte-Carlo
simulations, where edges terminating in polymer particles are assigned a much
lower conductivity. We count the number of steps that it takes each  particle to
reach one of the exterior vertices, repeating the random walk for each starting
point a certain number of times. From the histograms of the number of steps we
predict the cumulative release profile of the simulated matrix tablet. These
predictions are in good qualitative agreement with experimental matrix tablets
formulated from powder mixtures of different polymer contents \cite{DCDS_B}. As
we considered only monodisperse packings, no distinction between topological and
geometrical metrics for random walks (defined in the next section) was
necessary. However, it is well possible that the powder particles of different
compounds have different sizes. This leads to the question whether there is a
relationship between the two metrics. 

\section{Topological and Geometrical Random Walks}\label{section:TopGeomRandWalk}
Let $V\subset[0,1]^3$ be the set of centers of a bidisperse random sphere
packing on the flat torus (i.e.~a cube with periodic boundary conditions), where
all spheres have radii $r_1<r_2$ with $m_1$, respectively $m_2$ spheres in each
class. Let vertices $\mathbf{x}$ and $\mathbf{y}$ be the centers of two spheres
with radii $r_\mathbf{x}$ and $r_\mathbf{y}$ (which are equal to either $r_1$ or
$r_2$), and fix a constant $\varepsilon > 0$. Then $\mathbf{x}$ and $\mathbf{y}$
are joined by an edge if
\begin{equation*}
|\mathbf{x}-\mathbf{y}| \le (1+\varepsilon) (r_\mathbf{x}+r_\mathbf{y}).
\end{equation*}
The resulting graph with fixed vertex set $V$ and variable edge set
$E_\varepsilon$ is denoted $G_\varepsilon$. Clearly if $\varepsilon_1<
\varepsilon_2$, then  $E_{\varepsilon_1} \subset E_{\varepsilon_2}$. Contact
graphs of monodisperse random sphere packings were already studied by Powell
\cite{Powell1980}, who investigated their degree distributions as a function of
$\varepsilon\gtrsim0$. The infimum $\varepsilon_*$ of all $\varepsilon>0$ such
that $G_\varepsilon$ is connected, is called the \textit{level of
connectedness}.

Let $V_\alpha$ and $V_\omega$ be subsets of $V$, chosen randomly among all
vertices $V$. We perform random walks on the graph
$G_\varepsilon=(V,E_\varepsilon)$ that start in $V_\alpha$ and terminate in
$V_\omega$. For simplicity, we will work with equal weights for all edges in
this paper (but see the Discussion in section \ref{section:Discussion}). For
such a random walk we count the steps (the \textit{topological metric}) and it's
euclidean length (i.e.~the sum of the lengths of all the edges in the walk, the
\textit{geometric metric}). To avoid confusion, the euclidean length is
\textit{not} the euclidean distance between the initial and terminal
points of the walk.

Let $\pi_1,\,\pi_{12}$ and $\pi_2$ be the relative frequencies of an edge
$e\in E_\varepsilon$ of $G_\varepsilon=(V,E_\varepsilon)$ to be of length
$\le(1+\varepsilon)2r_1,\,\le(1+\varepsilon)(r_1+r_2)$ and
$\le(1+\varepsilon)2r_2$, respectively. Then we expect a linear
relationship for the average euclidean length of a walk with $n$ steps
\begin{equation}\label{length_vs_number}
l(n) \asymp (2r_1\pi_1+(r_1+r_2)\pi_{12}+2r_2\pi_2)n =: a_{th}n,
\end{equation}
in the sense that the quotient of the two expressions approaches $1$ as $n\to\infty$. 

We use an implementation of the Lubachevsky-Stillinger protocol developed by the
Complex Materials Theory Group at Princeton University \cite{Skoge} with
periodic boundary conditions. This is to avoid boundary effects that play an
important role for small packing sizes and decrease the ``maximum'' random
packing fraction $\phi$ (the volume occupied by the spheres divided by the total
volume of the container, $\approx 0.63$ for monodisperse packings). Unless
stated otherwise, we fix the total number of spheres to be 400. For the volume
fraction $\rho$ of small sphere volume $V_1$ vs.~the total sphere volume
$V_1+V_2$ we have 
\begin{equation*}
\rho = \frac{V_1}{V_1+V_2} = \frac{1}{1+\xi\sigma^3},
\end{equation*}
where $\xi=\frac{m_2}{m_1}$ is the ratio of large to small spheres and $\sigma
= \frac{r_2}{r_1}$ is the ratio of the sphere radii.

\section{Results}\label{section:Results}
\begin{figure}[th]
\begin{center}
\includegraphics[width=80mm]{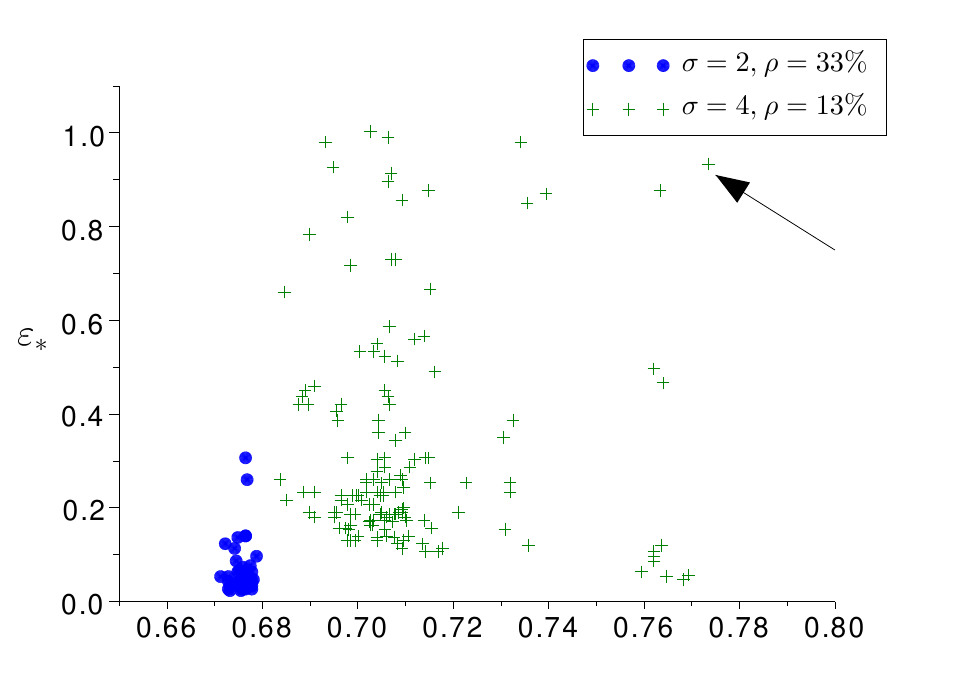}
\caption{Plot of the level of connectedness $\varepsilon_*$ vs.~the packing
fraction $\phi$ of $50$ (blue dots) respectively $150$ (red dots) packings for
two choices of aspect ratios $\sigma$ and volume fractions $\rho$. The radius
growth rates and termination criteria used in the Lubachevsky-Stillinger
protocol are identical. The arrow indicates an extreme case of high packing
fraction but poor level of connectedness. }\label{epsi_vs_phi}
\end{center}
\end{figure}

We consider first the correlation between the packing fraction $\phi$ and the
level of connectedness $\varepsilon_*$ of a packing. It is very hard to tightly
pack mixtures having a large aspect ratio $\sigma$ and a small volume fraction
$\rho$ of small spheres. Geometrically, an individual small sphere may float
freely in the interstitial space between the large spheres (a ``rattler''), even
though the overall packing fraction may be significantly higher than the packing
fraction of a monodisperse packing (see Figure \ref{epsi_vs_phi}). It is
possible to improve this by decreasing the sphere growth rate at high collision
rates or high pressures \cite{Biazzo}.

\begin{figure}[th]
\begin{center}
\includegraphics[width=90mm,height=60mm]{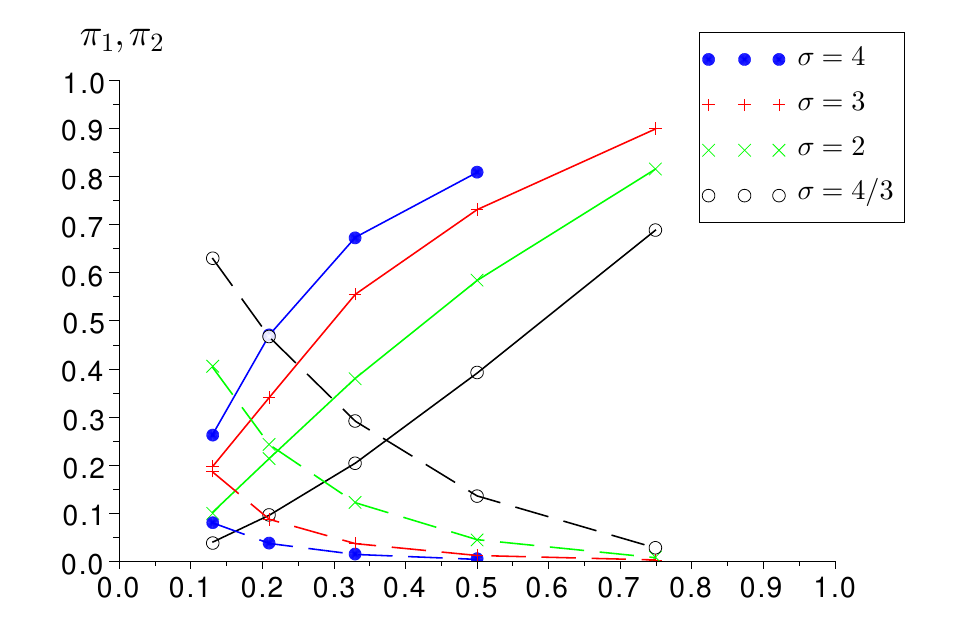}
\caption{Probabilities for an edge to be short ($\pi_1$, solid lines)
respectively long  ($\pi_2$, dashed lines) as functions of the volume fraction
$\rho$ of small spheres  and for different aspect ratios
$\sigma$.}\label{edge_distribution}
\end{center}
\end{figure}

After creating the sphere packings, we select those that have a level of
connectedness $\varepsilon_*\le 0.05$ for the further investigations. As
mentioned
before, their relative occurrence varies with the parameters of the sphere
mixture. In Figure \ref{edge_distribution} we show the probabilities $\pi_1$
and $\pi_2$ of short and long edges, respectively, as functions of 
$\sigma$ and  $\rho$.

\begin{figure}[th]
\begin{center}
\includegraphics[width=66mm]{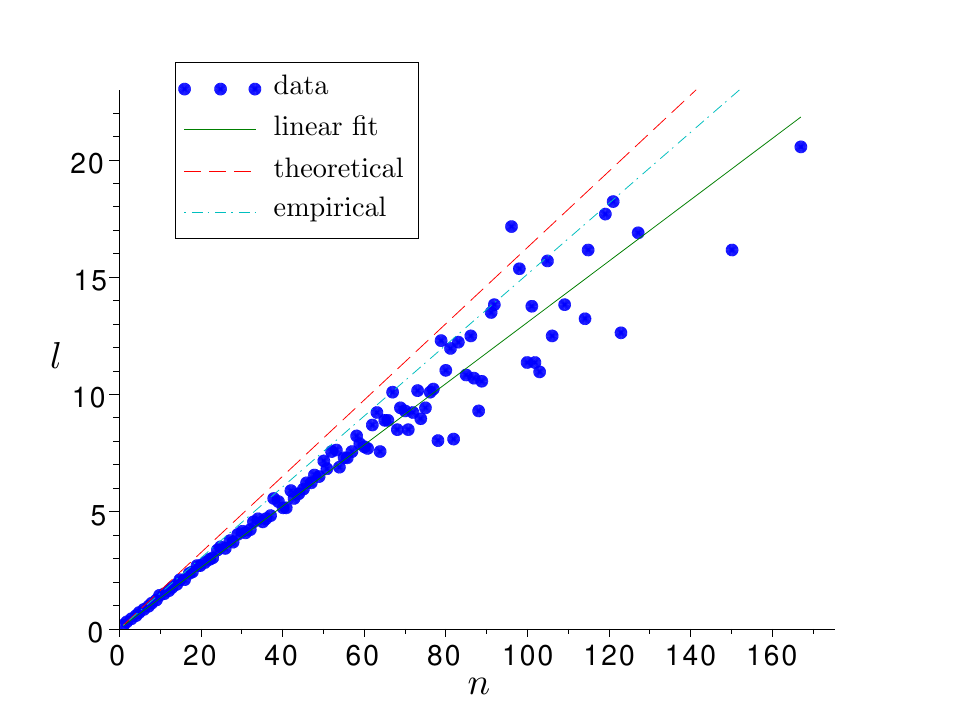}
\includegraphics[width=66mm]{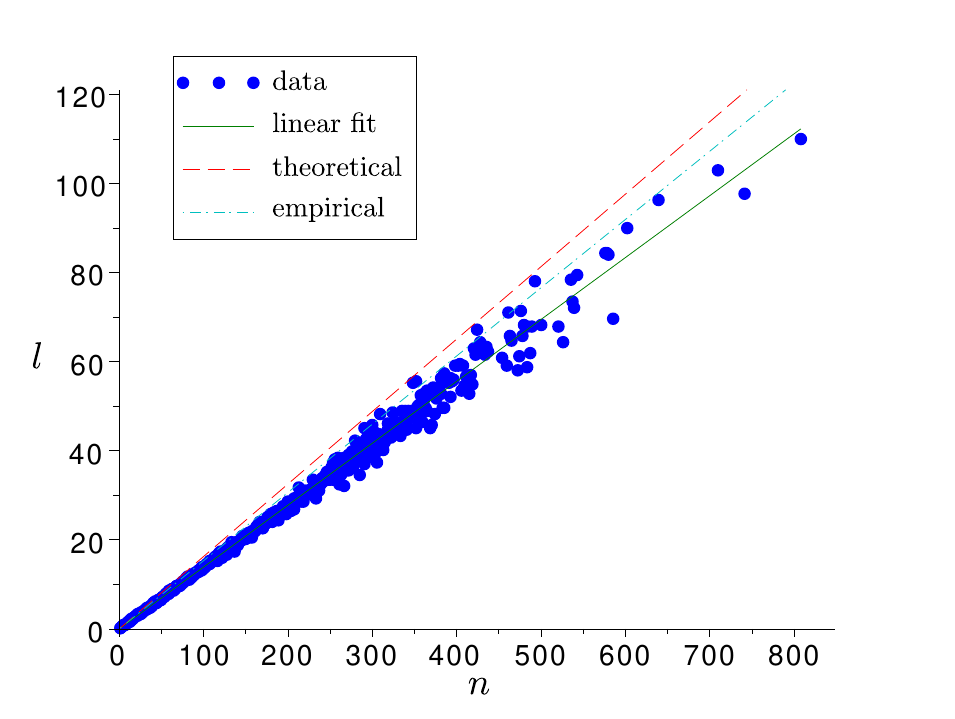}
\caption{\textit{(Left)} Plot of the average lengths of random walks of $n$
steps on a packing of 400 spheres with $\sigma =4$ and a volume fraction
$\rho=33\%$. A total of $10^4$ random walks were created, ending in a target set
of size $50$. The solid line shows a linear fit to the data (giving the slope
$a_{exp}$), while the dashed line has the slope predicted by equation
\eqref{length_vs_number}, where the relative frequencies  $\pi_1,\,\pi_{12}$ and
$\pi_2$ are taken from the contact graph. The dashed-dotted line uses the same
equation, but with the empirical edge length frequencies $\tilde{\pi}_1,\,
\tilde{\pi}_{12}$ and $\tilde{\pi}_2$ obtained from the random walks.
\textit{(Right)} As in the left panel, but with $|V_\omega|=10$.
}\label{linear_growth}
\end{center}
\end{figure}

\begin{figure}[th]
\begin{center}
\includegraphics[width=90mm]{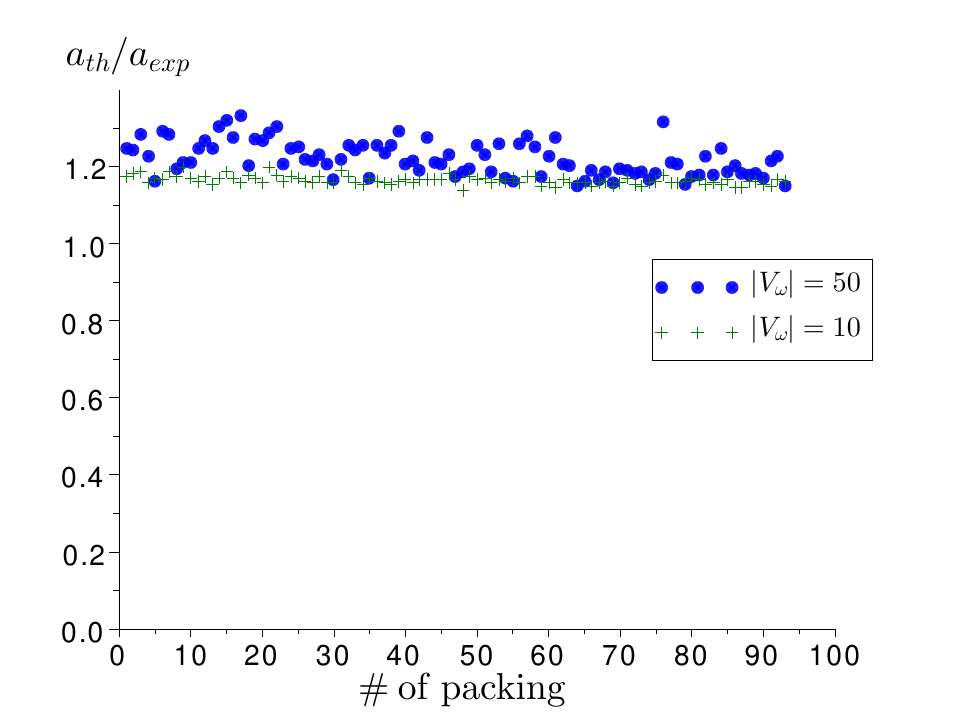}
\caption{A plot of the ratio $a_{th}/a_{exp}$ of theoretical and experimental
slopes for random walks on $93$ packings, with different sizes of the target
sets. 
}\label{ratios}
\end{center}
\end{figure}

For each packing we create $10^4$ unbiased random walks starting at a random
initial point and terminating in a random set of varying size. Each walk passes
through a certain number of short ($2r_1$), intermediate ($r_1+r_2$) and long
($2r_2$) edges. These numbers are recorded and divided by the total number of
steps in the walk giving the empirical edge length frequencies
$\tilde{\pi}_1,\,\tilde{\pi}_{12}$ and $\tilde{\pi}_2$. These can be compared to
the edge length frequencies  $\pi_1,\,\pi_{12}$ and $\pi_2$ obtained directly
from the graph that are used in \eqref{length_vs_number}.  We select all walks
that share the same number of steps $n$ and take the average of their euclidean
lengths, which gives the quantity $l(n)$. When we plot $l(n)$ against $n$, we
observe a linear relationship with slope $a_{exp}$ for small values of $n$,
before the points become scattered, see Figure \ref{linear_growth}. The
theoretical value $a_{th}$ from \eqref{length_vs_number} overestimates the
actual slope somewhat, but the ratio $a_{th}/a_{exp}$ between theoretical and
experimental slope decreases with the size of the target set (and hence an
increased average duration of the random walks), see Figure \ref{ratios}. 

\section{Discussion and Conclusion}\label{section:Discussion} 
The random walks considered here mimic the diffusion of a drug molecule from
its original position to the edge of a  matrix tablet, as was proposed in
\cite{DCDS_B}. Other applications of this paradigm are the diffusion of soluble
materials in porous soils, rocks or biological tissues. The
length distribution of the random walks is used to calculate the escape times
for molecules and the cumulative distribution function of the escape times is
taken as a prediction for the release profile of the tablet. If the particles of
the powder mixture have different sizes, the definition of the ``length'' of a
random walk needs to be revisited.

In this paper we have provided evidence that the geometric and topological
metrics on the
contact graph of a random dense sphere packing are related by the formula
\eqref{length_vs_number}, in the limit of long walks that sample correctly the
edge length distribution. While the average ratio $a_{th}/a_{exp}$ for the
walks on a packing of 400 spheres with size of the target set $|V_\omega|=10$ is
$1.16$ (Figure \ref{ratios}, ``+''), we expect this ratio to approach $1$ as the
number of spheres in the packing increases. This is the focus of ongoing
research.

Packings of hard spheres have found great interest in the computational physics
community, see \cite{Powell1980,Biazzo,Kamien,Song,Hermes,Farr}. Despite the
apparent simplicity of the concept, many quantities still remain to be
investigated, most prominently the maximum packing fraction of a random sphere
packing. Multicomponent mixtures are even more interesting because of multiple
possible edge lengths in the contact graph. Herein, we have determined the
frequencies $\pi_1$ and $\pi_2$ of short and long edges from contact graphs for
different types of sphere mixtures. As Figure \ref{edge_distribution} suggests,
these functions are monotone in both the aspect ratio $\sigma$ and the small
sphere volume fraction $\rho$ of the sphere mixture. We conjecture the
following limits of the edge length probabilities
\begin{equation*}
\begin{aligned}
\lim_{\sigma\to \infty} \pi_1(\sigma,\rho) &= \left\{\begin{array}{ll} 0
& \textrm{ if } \rho=0 \\ 1 & \textrm{ if } \rho >0 \end{array}\right., \\
\lim_{\sigma\to \infty} \pi_2(\sigma,\rho) &= \left\{\begin{array}{ll} 1
&
\textrm{ if } \rho=0 \\ 0 & \textrm{ if } \rho >0 \end{array}\right..
\end{aligned}
\end{equation*}
That is, for extremely large values of the ratio of large to small radii, short
edges are going to dominate, as soon as any small spheres  are present in
the mixture. It is difficult to make a corresponding prediction for
$\sigma\to 1^+$, as then the distinction between short, long and medium
edges disappears. Notice the change from concave to convex in $\pi_1$ as
$\sigma$ decreases from $4$ to $4/3$ in Figure
\ref{edge_distribution}, while no change of convexity is seen in the
corresponding values of $\pi_2$.

A natural generalization is to consider $k$-disperse packings with $k\ge 2$
radii classes. Then there are at most $\frac{k(k+1)}{2}$ possible edge lengths
in the contact graph and a straightforward generalization of equation
\eqref{length_vs_number} with the edge length expectation is likely to hold. The
situation is much more complicated if the radii of the spheres themselves (and
consequently the edge lengths) are a random variable.

In the present work we have considered only the case of random walks on a graph
where all edges carry the same weight. The motivating problem from
pharmaceutical science in \cite{DCDS_B} however required that some edges are
much harder to access than others (and in the extreme case, not at all), namely
those ending at particles of polymer
type. Although the edge length distribution $(\pi_1,\,\pi_{12},\,\pi_2)$ is the
same, a long random walk is then no longer going to sample all edges with these
frequencies. The conjectured length of random walks \eqref{length_vs_number}
will need to take that into account.

\section*{Acknowledgments }
I thank Monica Skoge and Aleksandar Donev of Princeton University for help with
the package \cite{Skoge} and three anonymous readers for valuable comments.

\bibliographystyle{elsarticle-num}
\bibliography{comp_phys}
\end{document}